\documentclass[aps,preprint]{revtex4}
\usepackage[dvips]{graphics,graphicx}
\begin{document}
\title{Master equation approach to conductivity of bosonic and fermionic carriers in one- and two-dimensional lattices }
\author{Andrey R. Kolovsky}
\affiliation{Kirensky Institute of Physics and Siberian Federal University, 660036 Krasnoyarsk, Russia} 
\date{\today}

\begin{abstract}
We discuss the master equation approach to diffusive current of bosonic or fermionic carriers in one- and two-dimensional lattices. This approach is shown to reproduce all known results of the linear response theory, including the integer quantum Hall effect for fermionic carriers. The main advantage of the approach is that it allows to calculate the current beyond the linear response regime where new effects are found. In particular, we show that the Hall current  can be inverted by changing orientation of the static force (electric field) relative to the primary axes of the lattice.

\end{abstract}

\maketitle

\section{Introduction}

Addressing conductance of a solid-state system one distinguishes two cases given by inequality relation between the mean free path $L_F=v_F \tau$ (here $v_F$ is the Fermi velocity and $\tau$ the collision time) and the system size $L$. If $L<L_F$ the conductance is due to  the ballistic transport and can be calculated using the Landauer-B\"uttiker formalism \cite{Butt85}. If $L\gg L_F$ the conductance is due to diffusive Ohmic current. In the past two decades the efforts were mainly aimed to the ballistic transport in mesoscopic solid-state systems \cite{Data95,Naza09}.  However, recently we have seen a recovery of interest to the diffusive transport, now with respect to the new experimental system -- ultracold atoms in optical lattices \cite{Ott04,69}.  Although cold atoms in optical lattices are equivalent to  electrons in a crystal, several features make this system different: (i) Optical lattices are free from defects. Therefore the collision time is defined only by the product of $s$-wave scattering length and atomic density, which can be varied at will \cite{Gust10}; (ii) Atoms are charge neutral. Because of this, to mimic the electric field,  experimentalists use the gravitational force \cite{Ande98}, gradient of the magnetic field \cite{Tarr12}, etc. When compared with solid-state systems these potential forces correspond to very large electric fields, far beyond the validity region of the linear response theory; (iii) Similar problem is faced for the synthetic magnetic field that is currently  realized by introducing the Peierls phase \cite{Aide11}. Again, when compared with crystal electron in a magnetic field,  this phase corresponds to extremely high magnetic flux density; (iv) The last but not the least, different species of atoms obey different quantum statistics.  All these features of the cold-atom system require critical revision of the theory of diffusive transport that is largely based on the linear response theory for fermionic carriers. 

In this work we revisit the master equation approach where one analyzes dynamics of the single-particle density matrix of the carriers. We will summarize our previous studies on bosonic conductivity \cite{77,86}  and complement them with new results on fermionic conductivity. To find the `current-voltage' characteristic we use in parallel `algebraic' and `dynamical' methods. In the algebraic approach we analytically or semi-analytically solve the equation for the stationary density matrix and then find the diffusive current.  Dynamical approach is straightforward numerical simulation of the system dynamics. We use it to check predictions of the algebraic method and to calculate the current for some specific system parameters (irrational $\beta$) where the algebraic method is not applicable. 

The paper consists of two parts, Sec.~\ref{sec2} and Sec.~\ref{sec3}, devoted to conductivity in one-dimensional and two-dimensional lattices, respectively.  The analysis is carried out in the tight-binding approximation. For two-dimensional lattices we consider the general case where both  `electric'  and `magnetic' fields are present (the so-called Hall configuration). We mention that the success of the algebraic method in treating this general case is mainly due to the recent progress in understanding the properties of the Landau-Stark states \cite{85,90} which, by definition, are eigenstates of a quantum particle in the Hall configuration.

\section{Diffusive current in one-dimensional lattices}
\label{sec2}

\subsection{The model}

Our theoretical framework  is the master equation for the single-particle density matrix of the carriers,
\begin{equation}
\label{1} 
\frac{d \hat{\rho}}{dt}=-\frac{i}{\hbar}[\widehat{H},\hat{\rho}] +{\cal L}(\hat{\rho}) \;,
\end{equation}
where $\widehat{H}$ is the carrier Hamiltonian in the tight-binding approximation,
\begin{equation}
\label{2} 
\widehat{H}=\widehat{H}_0 + Fd\sum_{l=1}^L
|l\rangle\langle l| \;,\quad \widehat{H}_0=-\frac{J}{2}\sum_{l=1}^L \left( |l+1\rangle l \langle l| +h.c.\right) \;,
\end{equation}
and ${\cal L}(\hat{\rho})$ the relaxation term. In the Hamiltonian (\ref{2}) $F$ is the magnitude of a static (electric) force, $d$ the lattice period, $J$ the hopping matrix element, and $L$ the number of lattice sites.  It is further assumed that for $F=0$ the system relaxes into the equilibrium state $\bar{\rho}_0$ and this process is characterized by the overall relaxation constant $\gamma$, i.e.,
\begin{equation}
\label{3} 
{\cal L}(\hat{\rho})=-\gamma(\hat{\rho} - \bar{\rho}_0) \;.
\end{equation}
The equilibrium  density matrix $\bar{\rho}_0$ is diagonal in the quasimomentum basis, 
\begin{equation}
\label{4} 
|k\rangle=L^{-1/2}\sum_l \exp(i2\pi kl/L) |l\rangle \;,
\end{equation}
and in the simplest case of zero temperature is given by
\begin{equation}
\label{5} 
\bar{\rho}_0= N |0\rangle\langle 0 |
\end{equation}
for bosonic carriers and  by 
\begin{equation}
\label{6} 
\bar{\rho}_0= \sum_{k=-N/2}^{N/2} |k\rangle\langle k | 
\end{equation}
for spinless fermions. In Eqs.~(\ref{5},\ref{6}) $N$ is the total number of carriers that determines the dimensionless density $n_B=N/L$ in the case of Bose particles and the Fermi energy $E_F$  in the case of Fermi particles. Our goal is to calculate the stationary current
\begin{equation}
\label{7} 
\bar{v}={\rm Tr}[\hat{v}\bar{\rho}] \;,
\end{equation}
where $\bar{\rho}$ is the stationary solution of the master equation (\ref{1}-\ref{3}) and  $\hat{v}$ the current operator
\begin{equation}
\label{8} 
\hat{v}=\frac{v_0}{2i}\sum_l \left(|l+1\rangle\langle l| -h.c.\right) \;,\quad v_0=\frac{dJ}{\hbar} \;.
\end{equation}

Since the trace of a matrix is invariant with respect to unitary transformations we can use any complete basis to evaluate Eq.~(\ref{7}). Two natural choices are the basis of Bloch states (\ref{4}), which are eigenstates of the Hamiltonian $\widehat{H}_0$,  and the basis of Wannier-Stark states $|n\rangle$,
\begin{equation}
\label{9} 
|n\rangle=\sum_l {\cal J}_{l-n}\left(\frac{J}{2F}\right) |l\rangle,
\end{equation}
which are eigenstates of the Hamiltonian $\widehat{H}$.  It appears that calculations are easier in the Wannier-Stark basis. In fact, using this basis one immediately finds the stationary density matrix,
\begin{equation}
\label{10} 
\bar{\rho}(n,n')=\frac{\hbar\gamma}{\hbar\gamma+i(E_{n'}-E_n)} \bar{\rho}_0(n,n')  \;, 
\end{equation}
where $E_n=Fdn$ is the spectrum of the Wannier-Stark states. The explicit form of the stationary matrix in the Bloch basis is not so obvious. However,  the Bloch basis is more attractive from the viewpoint of  physical interpretation because in this basis the current operator is a diagonal matrix
\begin{equation}
\label{11} 
\langle \kappa'|\hat{v}| \kappa\rangle = v(k)\delta(\kappa'-\kappa)  \;, \quad v(\kappa)=v_0 \sin \kappa  \;.
\end{equation}
(From now on we assume the limit $L\rightarrow\infty$, where the quasimomentum $\kappa=2\pi k/L$ is continuous quantity.) Thus we need to know only diagonal elements of the stationary density matrix (\ref{10}) in the Bloch basis,  which are interpreted as populations of the quasimomentum states.  We will discuss these two approaches in more detail in the next subsection where, to simplify equations, we set the lattice period $d$ and Planck's constant $\hbar$ to unity.

\subsection{Bosonic conductivity}

First we consider the case of bosonic carriers. Using the Wannier-Stark basis (\ref{9}) the stationary current (\ref{7}) was calculated in Ref.~\cite{Mino04}.  The resulting equation was proved to exactly reproduce the Esaki-Tsu equation for the diffusive current,
\begin{equation}
\label{a1} 
\frac{\bar{v}}{v_0}=n_B  \frac{F/\gamma}{1+(F/\gamma)^2} \;,
\end{equation}
which was introduced by Esaki and Tsu in 1970 for the semiconductor super-lattices \cite{Esak70}.  In the parameter region $F>\gamma$ (which physically means that the  Bloch frequency $\omega_B=Fd/\hbar$ exceeds the inverse relaxation time $\gamma/\hbar$) Eq.~(\ref{a1}) describes the phenomenon of negative differential conductivity, where the current decreases with increase of the voltage. The opposite limit $F \ll \gamma$ corresponds to the linear response regime, where the current is proportional to $F$. 

Using the Bloch basis the current (\ref{7}) and stationary velocity distribution of the carriers were calculated in Ref.~\cite{77}. The key point of Ref.~\cite{77}  was to present the relaxation operator (\ref{3}) in the Lindblad form \cite{Lind76}.  For zero temperature this can be done exactly,  resulting in the following equation for diagonal elements of the stationary density matrix:
\begin{equation}
\label{a2}
-F\frac{\partial f(\kappa)}{\partial \kappa} -\gamma f(\kappa)=-\gamma\delta(\kappa) \;,\quad 
f(\kappa)=\langle \kappa|\bar{\rho}| \kappa\rangle \;.
\end{equation}
Notice that in (\ref{a2})  the density matrix is normalized to unity, hence, the mean current is given by
\begin{equation}
\label{a3} 
\frac{\bar{v}}{v_0}=n_B  \int_{-\pi}^\pi \sin\kappa \; f(\kappa) {\rm d} \kappa \;, 
\end{equation}
where $n_B$ is the carrier density. To make the paper self-consistent we present an alternative derivation of Eq.~(\ref{a2}) in the next paragraph. 

Let us denote by $\bar{\rho}(\kappa',\kappa)$ matrix elements of the stationary density matrix and by $H(\kappa',\kappa)$ matrix elements of the Hamiltonian (\ref{2}).  To calculate $H(\kappa',\kappa)$ we use the identity $\sum_n \exp(i\kappa n) |n\rangle= |\kappa\rangle$, where $|\kappa\rangle$ are the Bloch states (\ref{4}) and $| n\rangle$ the Wannier-Stark states (\ref{9}). Then the matrix elements are  
\begin{equation}
\label{a4} 
H(\kappa',\kappa)=\frac{1}{L}\sum_n n e^{i(\kappa'-\kappa)n}
= -iF\frac{\partial}{\partial \kappa'}\delta(\kappa'-\kappa)
= iF\frac{\partial}{\partial \kappa}\delta(\kappa'-\kappa) \;.
\end{equation}
Using (\ref{a4}) the equation  
for the stationary matrix takes the form
\begin{equation}
\label{a5}
-F \left(\frac{\partial \bar{\rho}(\kappa',\kappa)}{\partial \kappa'}
+\frac{\partial \bar{\rho}(\kappa',\kappa)}{\partial \kappa}\right)
-\gamma \bar{\rho}(\kappa',\kappa)=-\gamma \delta(\kappa')\delta(\kappa) \;.
\end{equation}
Finally, considering the density matrix as the function of $\xi=(\kappa'+\kappa)/2$ and $\eta=(\kappa'-\kappa)/2$, we obtain Eq.~(\ref{a2}) for the diagonal elements $f(\kappa)=\bar{\rho}(\xi=\kappa,\eta=0)$.

\begin{figure}
\center
\includegraphics[width=12cm, clip]{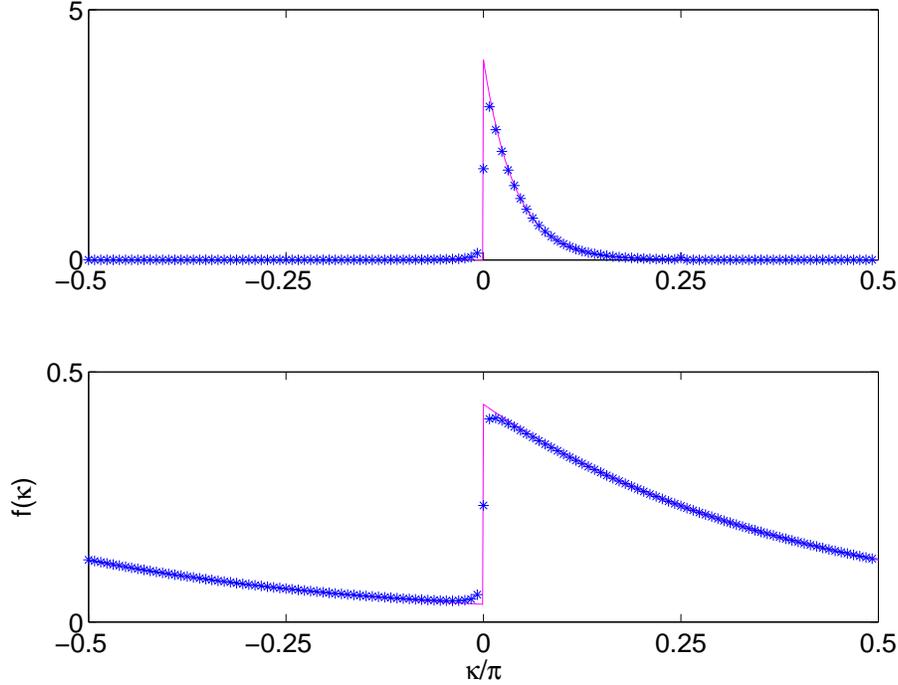}
\caption{Stationary distributions $f(\kappa)$ of bosonic carries over the quasimomentum states. The solid lines and asterisks are analytical and numerical results, respectively.  Parameters are $J=1$, $\gamma=0.4$,  $L=128$, and $F=0.1$ (upper panel) and $F=1$ (lower panel, notice different scale for the $y$ axis).}
\label{fig1}
\end{figure}

With the quasimomentum $\kappa$ in the interval $0< \kappa <2\pi$ the solution of Eq.~(\ref{a2}) reads
\begin{equation}
\label{a6}
f(\kappa)=\frac{\gamma}{F}\frac{1}{1-\exp(-2\pi\gamma/F)}
\exp\left(-\frac{\gamma}{F}\kappa \right) \;. 
\end{equation}
Thus, when we consider $\kappa$ within the first Brillouin zone $-\pi\le\kappa<\pi$, the function  $\bar{\rho}(k)$ has a jump at $\kappa=0$ (see solid lines in Fig.~\ref{fig1}). If $F\ll \gamma$ Eq.~(\ref{a6}) simplifies to
\begin{equation}
\label{a7}
f(\kappa)=\left\{
\begin{array}{l}
\frac{\gamma}{F}\exp\left(-\frac{\gamma}{F}\kappa \right) \quad {\rm for} \quad \kappa>0 \\
0 \quad {\rm for} \quad \kappa<0
\end{array}
\right.  \;,
\end{equation}
that has a simple physical interpretation --  finite $F$ smoothes the $\delta$-peaked distribution of degenerate bosons into exponential function with its tail extended towards positive or negative $\kappa$ depending on the sign of $F$.  Substituting Eq.~(\ref{a7}) into Eq.~(\ref{a3}) we obtain
\begin{equation}
\label{a8} 
 \frac{\bar{v}}{v_0}\approx  
 n_B  \int \kappa \frac{\gamma}{F}\exp\left(-\frac{\gamma}{F}\kappa \right) {\rm d} \kappa 
=n_B \frac{F}{\gamma} \;.
\end{equation}
It is also easy to show that by using the exact distribution (\ref{a6}) instead of the approximate distribution (\ref{a7}) we recover the Esaki-Tsu equation (\ref{a1}) that describes the diffusive current beyond the linear response regime.

It is interesting to compare the analytical results (\ref{a6}-\ref{a7}) against numerical simulations of the system dynamics. In our numerical approach we solve the evolution equation (\ref{1}) in the Wannier basis for the initial condition given by the equilibrium density matrix (\ref{5}). For this initial condition the carriers show decaying Bloch oscillations with the Bloch frequency $\omega_B=dF/\hbar$ and decay time $\tau=2\pi\hbar/\gamma$. After a few oscillations the current stabilizes at its stationary value $\bar{v}$.  When this steady state is reached, we extract $f(\kappa)$ by Fourier transforming the density matrix.  As an example, Fig.~\ref{fig1} shows stationary distributions for $J=1$, $\gamma=0.4$,  and $F=0.1$ (upper panel, linear response regime) and $F=1$ (lower panel, strong forcing beyond the linear response regime). Numerical results are seen to nicely reproduce the dependence (\ref{a6}), where small deviations are due to finite size of the system used in numerical simulations.

\subsection{Fermionic conductivity}

\begin{figure}
\center
\includegraphics[width=12cm, clip]{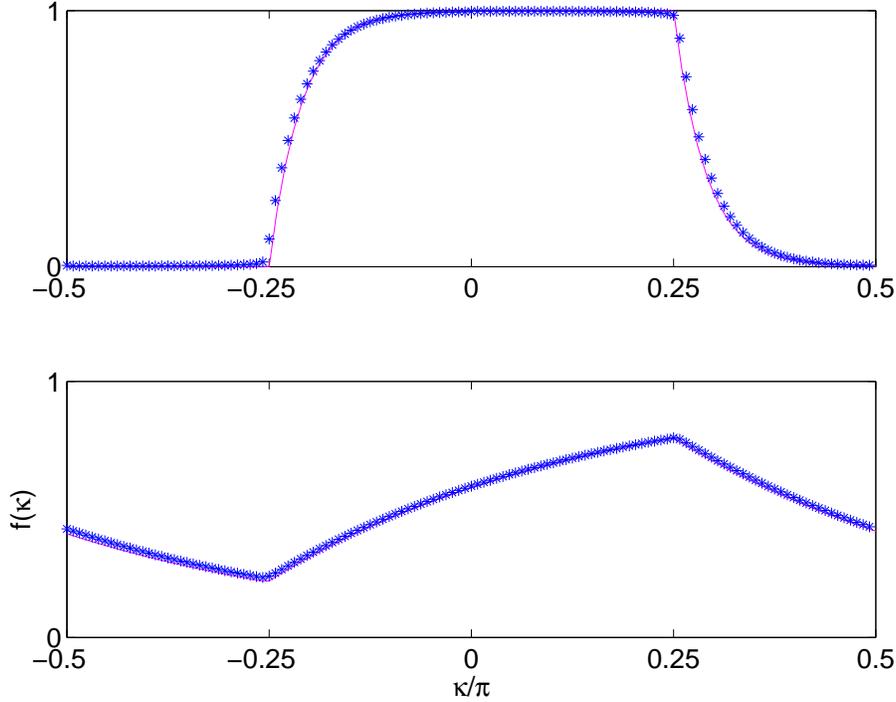}
\caption{The same as in Fig.~\ref{fig1} yet for fermionic carriers. The Fermi energy is set to zero, i.e., $\kappa_F=\pi/4$.} 
\label{fig2}
\end{figure}

In the case of fermionic carriers Eq.~(\ref{a6}) for diagonal elements of the stationary density matrix takes the form 
\begin{equation}
\label{b1}
-F\frac{\partial f(\kappa)}{\partial \kappa} -\gamma f(\kappa)=-\gamma f_0(\kappa) \;,
\end{equation}
where 
\begin{equation}
\label{b2}
f_0(\kappa)=\left\{
\begin{array}{l}
1 \quad {\rm for} \quad |\kappa| \le \kappa_F \\
0 \quad {\rm for} \quad |\kappa| >\kappa_F
\end{array}
\right. 
\end{equation}
is the zero temperature Fermi distribution. (Notice that now we do not normalize the density matrix to unity.) Considering the quasimomentum $\kappa$ in the interval  $0\le \kappa <2\pi$ the solution of (\ref{b1}-\ref{b2}) reads
\begin{equation}
\label{b3}
f(\kappa)=\left\{
\begin{array}{l}
1-a\exp\left[-\frac{\gamma}{F}\kappa \right]   \quad {\rm for} \quad 0\le \kappa <2\kappa_F \\
b\exp\left[-\frac{\gamma}{F}(\kappa-2\kappa_F)\right]      \quad {\rm for} \quad  2\kappa_F\le \kappa <2\pi 
\end{array}
\right.  \;,
\end{equation}
where coefficients $a$ and $b$ satisfy the following algebraic equation:
\begin{equation}
\label{b4}
\left(
\begin{array}{cc}
1&\exp[-2\gamma \kappa_F/F] \\
\exp[-\gamma(1-2\kappa_F)/F]&1
\end{array}
\right)
\left(
\begin{array}{c}   a\\b \end{array}
\right)
=\left(
\begin{array}{c}  1\\1 \end{array}
\right) \;.
\end{equation}
The exact solution (\ref{b3}-\ref{b4})  simplifies in the limit of small $F$ where the stationary distribution only slightly deviates from the Fermi distribution (\ref{b2}). We have
\begin{equation}
\label{b5} 
f(\kappa)=\exp\left[-\frac{\gamma}{F}(\kappa-\kappa_F)\right] \;, \quad \kappa> \kappa_F \;,
\end{equation}
for the right edge of the Fermi surface and 
\begin{equation}
\label{b6} 
f(\kappa)=1-\exp\left[-\frac{\gamma}{F}(\kappa-\kappa_F)\right] \;, \quad \kappa> -\kappa_F \;,
\end{equation}
for the left edge. Examples of the above distributions are given in Fig.~\ref{fig2} together with results of the dynamical approach described in the previous subsection.

To find the current  we should integrate the stationary distribution with the weight $v(\kappa) =v_0 \sin \kappa$. Because $v(\kappa)$ is an antisymmetric function of $\kappa$,  the integral reduces to 
\begin{equation}
\label{b7} 
\bar{v}= v_0 \int \sin \kappa \; \exp\left(-\frac{\gamma}{F}|\kappa-\kappa_F|\right) {\rm d} \kappa 
\approx 2v_F\frac{F}{\gamma} \;,
\end{equation}
where we explicitly assume the limit of small $F$ and $v_F=v_0\sin \kappa_F$ is the Fermi velocity. 
Since $k_F$ is uniquely defined by the density $n_F=N/L\le 1$ of the fermionic carriers, Eq.~(\ref{b7}) is the fermionic analogue of  the bosonic  equation (\ref{a8}). Moreover, it can be proved that the exact functional dependence of the current on $F$ is again given by the Esaki-Tsu equation,
\begin{displaymath}
\frac{\bar{v}}{v_0}=A(n_F) \frac{F/\gamma}{1+(F/\gamma)^2} \;,
\end{displaymath}
where the prefactor $A(n_F)$ takes values between 0 and 2 depending on the  filling of the ground Bloch band.

Concluding this section we briefly discuss two complementary viewpoints on the transport phenomena in the case of degenerate fermionic carriers \cite{remark}. The obtained approximate Eq.~(\ref{b7}) justifies the viewpoint that only the carriers which are at the Fermi surface take part in the transport. The other viewpoint is that all carriers below the Fermi surface participate in the transport, however, the current in opposite directions is largely compensated. In what follows we adopt the second point of view. As it will be shown in Sec.~\ref{sec3}, it provides a natural explanation for the integer quantum Hall effect in infinite two-dimensional lattices.

%
\section{Diffusive Hall current in two-dimensional lattices} 
\label{sec3}

\subsection{The model}

Results of the previous section prove that the master equation approach correctly describes the diffusive Ohm current in one-dimensional lattices. In this section we use it to analyze the Hall and Ohm currents in two-dimensional lattices. Our model Hamiltonian reads
\begin{equation}
\label{c1}
\widehat{H}= \widehat{H}_0 +\sum_{l,m} |l,m\rangle (F_x l+F_y m) \langle l,m |  \;,
\end{equation}
\begin{eqnarray}
\label{c2}
\widehat{H}_0= -\frac{J_x}{2} \sum_{l,m} \left(|l+1,m\rangle \langle l,m | e^{i2\pi\alpha m} + h.c.\right) 
-\frac{J_y}{2} \sum_{l,m} \left(|l,m+1\rangle \langle l,m |  + h.c.\right)  \;,
\end{eqnarray}
where $J_x$ and $J_y$ are the tunneling rates in the $x$ (index $l$) and $y$ (index $m$) directions,  $F_x$ and $F_y$ are two components of the static force ${\bf F}$, and $\alpha$  is the Peierls phase ($|\alpha|\le1/2$) that  accounts for  the real (charged particles) or artificial (charge neutral particles) magnetic field.   As before, the problem is to calculate the stationary currents
\begin{equation} 
\label{c3}
v_{x}={\rm Tr}[\hat{v}_{x}\bar{\rho}] \;, \quad v_{y}={\rm Tr}[\hat{v}_{y}\bar{\rho}] \;, 
\end{equation}
where $\hat{v}_{x}$ and $\hat{v}_y$ are the current operators,
\begin{eqnarray}
\label{c4}
\hat{v}_x=\frac{v_0}{2i}\sum_{l,m}\left(|l+1,m\rangle\langle l,m| e^{i2\pi\alpha m} -h.c.\right) \;,\quad 
\hat{v}_y=\frac{v_0}{2i}\sum_{l,m} \left( |l,m+1\rangle\langle l,m| -h.c.\right) \;,
\end{eqnarray}
and $\bar{\rho}$ is the stationary solution of the master equation (\ref{1}) with the relaxation term (\ref{3}). In the case of fermionic  carriers the equilibrium density matrix $\bar{\rho}_0$ in Eq.~(\ref{3}) is obviously given by 
\begin{equation}
\label{c5}
\bar{\rho}_0=\sum_{j=1}^{\cal N} |\Phi_j\rangle\langle \Phi_j| \;,
\end{equation}
where the sum includes all energy states $|\Phi_j\rangle$ of the Hamiltonian $\widehat{H}_0$ below the Fermi energy. The case of bosonic carriers is not so obvious because of possible degeneracy of the ground state. For this reason we focus from now on the case of fermionic carriers.
\begin{figure}
\center
\includegraphics[width=12cm, clip]{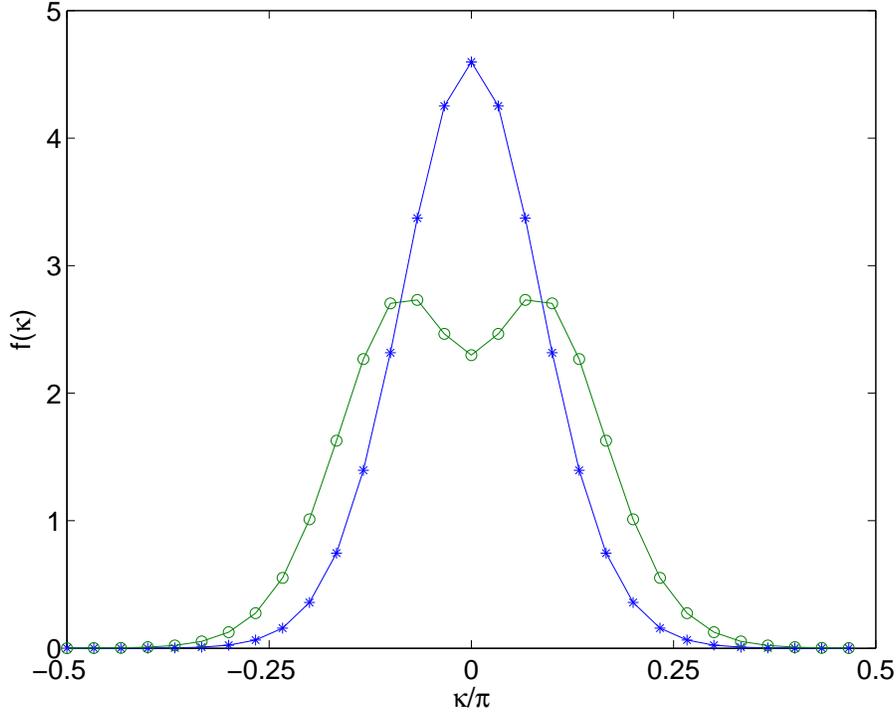}
\caption{Equilibrium distributions of the fermionic carriers for $\alpha=0.1$ and $J_x=J_y=1$. The Fermi energy is chosen in the energy gap above the first (asterisks) and the second (open circles line) magnetic bands.} 
\label{fig3}
\end{figure}

It is instructive to discuss the velocity distribution of the carriers in the equilibrium state (\ref{c5}). First of all we note that the velocity operators (\ref{c4}) do not commute if $\alpha\ne0$. Thus two-dimensional distribution function $f=f(v_x,v_y)$ is not defined and one has to deal with one-dimensional `reduced' distributions $f_x=f_x(v_x)$  and $f_y=f_y(v_x)$.  These distributions are given by diagonal elements of the matrix (\ref{c5}) in the basis of the operators  $\hat{v}_x$ and  $\hat{v}_y$, respectively. It is easy to prove that eigenstates of the velocity operators are Bloch waves in $x$ or $y$  directions:
\begin{equation}
\label{c6}
|\psi(\kappa_x,m_0)\rangle=\sum_{l,m}\frac{e^{i\kappa_x l}}{\sqrt{L}} \delta_{m,m_0}e^{i2\pi\alpha m} |l,m\rangle \;,\quad
|\psi(\kappa_y,l_0)\rangle=\sum_{l,m}\frac{e^{i\kappa_y m}}{\sqrt{L}} \delta_{l,l_0}  |l,m\rangle \;.
\end{equation}
Thus, similar to the case of 1D lattices, we can consider distributions $f_x$ and $f_y$ as the functions of the quasimomentum, where the velocity is related to the quasimomentum by the sine dispersion relation. Therefore the mean current is given by the equation
\begin{equation} 
\label{c7}
\frac{v_x}{v_0}= \int_{-\pi}^{\pi} \sin \kappa_x \; f_x(\kappa_x) {\rm d}\kappa_x  \;,
\end{equation}
and we have a similar expression for $v_y$.  The equilibrium distribution $f_x(\kappa_x)$ of the fermionic carriers is shown in Fig.~\ref{fig3} for $\alpha=1/10$,  where the energy spectrum of the system (\ref{c2})  consists of 10 magnetic bands. Two curves in Fig.~\ref{fig3} correspond to the Fermi energy within the first and the second energy gaps.  The equilibrium distribution $f_y(\kappa_y)$ obviously coincide with $f_x(\kappa_x)$ due to the gauge invariance of the problem. We also mention that, unlike the previously considered case of 1D lattices, here we cannot introduce the notion of Fermi wave vector because  $f_x(\kappa_x)$ are $f_y(\kappa_y)$ are smooth functions of the quasimomentum even at zero temperature.

\subsection{Landau-Stark states}
\label{sec3b}

We proceed with diffusive current for $F\ne0$. It was mentioned in Sec.~\ref{sec2} that calculation of the diffusive current in one-dimensional lattices is easier in the basis of the Hamiltonian $\widehat{H}$ which are the Wannier-Stark states. Similarly, to evaluate Eq.~(\ref{c3}) for the diffusive current in two-dimensional lattices it is convenient to use eigenstates of the Hamiltonian (\ref{c1}) which are termed the Landau-Stark states. Going ahead we mention that the spectrum and localization properties  of the Landau-Stark states crucially depend on the parameter $\beta=F_x/F_y$. We postpone the discussion of this issue to Sec.~\ref{sec3c} and focus on the simplest case $\beta=0$.

To find Landau-Stark states for $\beta=0$ one can use the same ansatz that is used to find the Landau states, namely,
\begin{equation}
\label{d1}
|\Psi \rangle= \sum_{l,m}\frac{e^{i\kappa l}}{\sqrt{L}} b_m  |l,m\rangle \;.
\end{equation}
This ansatz results in the following equation for the coefficients $b_m$,
\begin{equation}
\label{d2}
-\frac{J_y}{2}(b_{m+1}+b_{m-1}) + [F_y m - J_x\cos(2\pi\alpha m-\kappa)]b_m =E b_m \;,
\end{equation}
which is the Harper equation complimented with the Stark term. It follows from Eq.~(\ref{d2}) that the spectrum of Landau-Stark states consists of infinite number of the energy bands $E_n(\kappa)$, where asymptotically  
\begin{equation}
\label{d3}
E_n(\kappa)=F n - J_x\cos(\kappa-2\pi \alpha n)  \;, \quad F\equiv F_y\gg J_x \;.
\end{equation}
It also follows from Eqs.~(\ref{d1}-\ref{d2}) that the Landau-Stark states are localized functions in the direction parallel to the vector ${\bf F}$ and extended functions in the orthogonal direction. Another important feature of the Landau-Stark states is that they are transporting states, i.e., 
\begin{equation}
\label{d4} 
\langle \Psi_{n,\kappa} |\hat{v}_x | \Psi_{n,\kappa} \rangle \ne 0 \;. 
\end{equation}
Moreover, if $|\alpha|\ll 1/2$ and $F<F_{cr}=2\pi\alpha J_x$   there is a subset of states for which the quantity (\ref{d4})  equals to the drift velocity of the classical particle, 
\begin{equation}
\label{d5} 
v^*=F/2\pi \alpha \;. 
\end{equation}
As shown in Ref.~\cite{85}, these states are responsible for the ballistic Hall current  if the relaxation constant $\gamma=0$.

In the basis of the Landau-Stark states the stationary density matrix reads
\begin{equation}
\label{d6} 
\bar{\rho}(n,\kappa;n',\kappa')=\frac{\gamma}{\gamma+i[E_n'(\kappa')-E_n(\kappa)]} \bar{\rho}_0(n,\kappa;n'\kappa') \;.
\end{equation}
Equation (\ref{d6}) is the two-dimensional analogue of Eq.~(\ref{10}). Substituting Eq.~(\ref{d6}) into Eq.~(\ref{c3}) we calculate the diffusive Ohm ($v_y$) and Hall ($v_x$) currents. We also mention that the numerical calculations can be greatly simplified due to the fact that  the current operators (\ref{c4}) are diagonal matrices in the Landau-Stark basis with respect to the quasimomentum $\kappa$, i.e., $\langle \Psi_{n',\kappa'} |\hat{v}_{x,y} | \Psi_{n,\kappa} \rangle=v_{n',n}^{(x,y)}\delta(\kappa'-\kappa)$. Thus, instead of using the full $L^2\times L^2$ density matrix (\ref{d6}), we may use the $\kappa$-specific density matrix ${\cal R}_{n,n'}(\kappa)=\bar{\rho}(n,\kappa;n';\kappa)$ of the size  $L\times L$  \cite{86}.

\subsection{Diffusive Hall and Ohm currents}
\label{sec3c}

\begin{figure}[b]
\center
\includegraphics[width=12cm, clip]{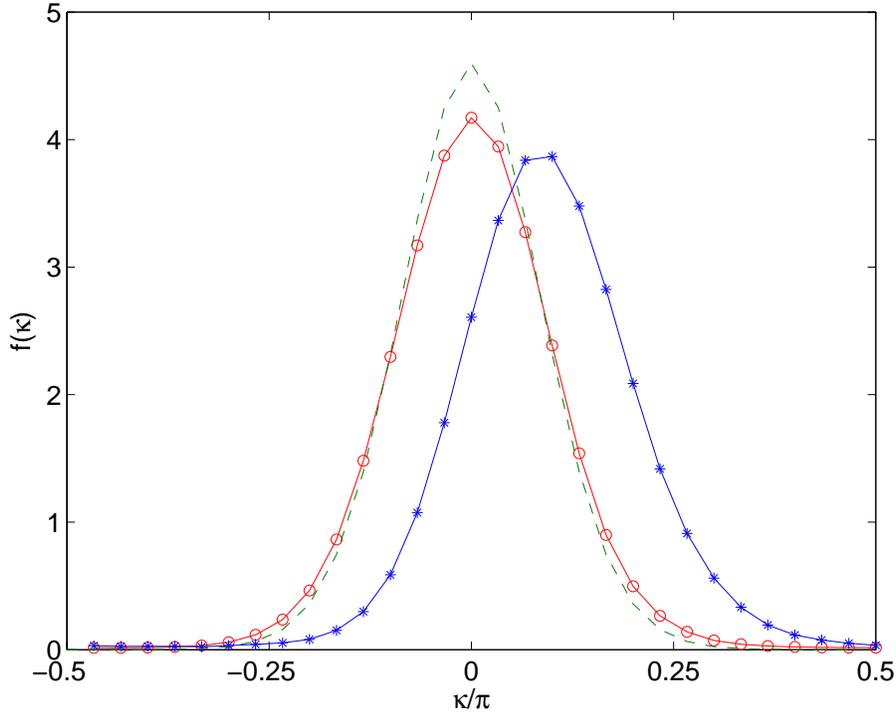}
\caption{The velocity distribution $f_x=f_x(\kappa_x)$, asterisks, and  $f_y=f_y(\kappa_y)$, circles, of the fermionic carriers for $F=0.2$. The other parameters are $J_x=J_y=1$, $\alpha=0.1$, $\gamma=0.1$, and  $\beta=F_x/F_y=0$. The dashed line corresponds to the equilibrium distributions when $F=0$.} 
\label{fig4}
\end{figure}
\begin{figure}
\center
\includegraphics[width=12cm, clip]{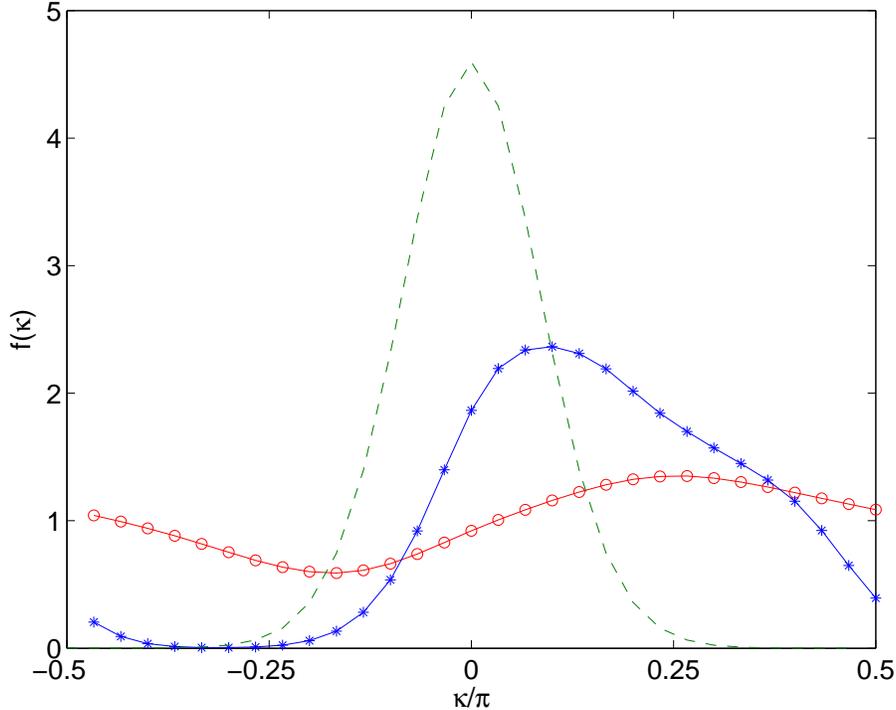}
\caption{The same as in Fig.~\ref{fig4} yet $F=1$.} 
\label{fig5}
\end{figure}

Figure \ref{fig4} shows the stationary velocity distributions of the carriers for $\gamma=0.1$ and $F=0.2$. These distributions should be compared with the dashed line which shows equilibrium distributions for $F=0$. It is seen that the applied force shifts distributions towards positive velocities and breaks the symmetry between  $f_x$ and $f_y$. This is consistent with results of the linear response theory,
\begin{equation}
\label{d7}
\frac{\bar{v}_{x}}{v_0}=\frac{F}{\gamma}\frac{\omega_c/\gamma}{1+(\omega_c/\gamma)^2} \;,\quad
\frac{\bar{v}_{y}}{v_0}=\frac{F}{\gamma}\frac{1}{1+(\omega_c/\gamma)^2} \;.
\end{equation}
where $\omega_c$ is the cyclotron frequency that for the considered lattice model is given by 
\begin{equation}
\label{d8} 
\omega_c=2\pi\alpha\sqrt{J_x J_y}/\hbar  \;. 
\end{equation}
It follows from Eqs.~(\ref{d7}-\ref{d8}) that the Hall current is mainly determined by the magnetic field and it is finite even in the limit $\gamma\rightarrow0$. On the contrary, the Ohm current is mainly determined by relaxation processes and vanishes if $\gamma=0$. 

\begin{figure}
\center
\includegraphics[width=12cm, clip]{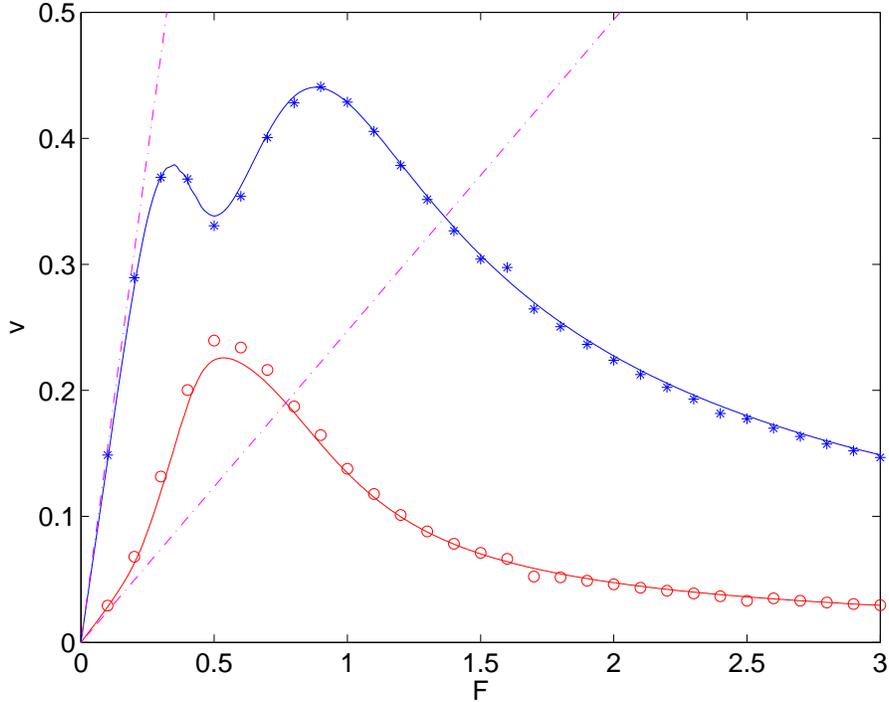}
\caption{Diffusive Hall (upper curve) and Ohm (lower curve) currents as functions of the static field $F$. The dashed lines are linear response equations (\ref{d7}). Symbols show results of dynamical approach for a small lattice comprising $20\times20$ sites.} 
\label{fig6}
\end{figure}

By further inspection of Fig.~\ref{fig4} we notice that the velocity distributions for $F\ne0$ are broader than those for $F=0$. This broadening is the origin of deviations from the linear response theory. As an example, Fig.~\ref{fig5} shows the distribution functions $f_x(\kappa_x)$ and $f_y(\kappa_y)$ for $F=1$, well beyond the linear response regime. Given the velocity distributions we can calculate the mean current from Eq.~(\ref{c7}). Alternatively, we can employ the method of Ref.~\cite{86} that allows to use much finer discretization of $\kappa$ and, hence, provides more accurate results. The Ohm and Hall currents are depicted in Fig.~\ref{fig6} together with predictions of the linear response theory (\ref{d7}). The dependence $\bar{v}_x=\bar{v}_x(F)$  for the Ohm current is seen to qualitatively reproduce the Esaki-Tsu equation (\ref{a1})  for the diffusive current in one-dimensional lattices. The dependence $\bar{v}_y=\bar{v}_y(F)$ for the Hall current, however, is not related to this equation -- a seeming similarity is accidental and does not appear for other system parameters, see Fig.~\ref{fig9} below. The particular form of the function $\bar{v}_y=\bar{v}_y(F)$ is determined by a sophisticated  interplay between the Bloch and cyclotron oscillations that is encoded in the Landau-Stark states. 

\subsection{Integer quantum Hall effect}
\label{sec3d}

In the previous subsection we analyzed the Hall and Ohm currents as functions of the electric field $F$. It is interesting to study  the currents as functions of the magnetic field that in the tight-binding approximation is characterized by the Peierls phase $\alpha$. It is expected that the Hall current should show a step-like behavior -- the phenomenon known as the integer quantum Hall effect. 

\begin{figure}
\center
\includegraphics[width=14cm, clip]{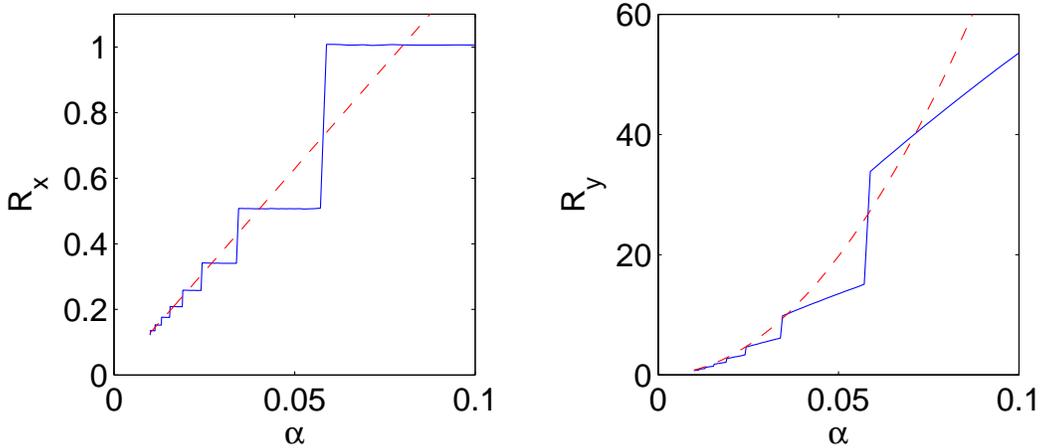}
\caption{Resistance $R_{x}=F_y/\bar{v}_{x}$ (left panel) and  $R_{y}=F_y/\bar{v}_{y}$ (right panel) as functions of the Peierls phase. The system parameters are $J_x=J_y=1$, $\gamma=0.01$, $F_y=0.01$, and  $E_F=-1.5$ (should be compared with $E=-2$ that is the bottom of the Bloch band for $\alpha=0$). The dashed lines show resistances calculated by using Eqs.~(\ref{a1}).} 
\label{fig7}
\end{figure}

Numerical result presented in Fig.~\ref{fig7}  confirms that the master equation approach fairly reproduces this effect.  Figure \ref{fig7} shows the Hall (left panel) and Ohm (right panel) resistance $R_{x,y}=F_y/\bar{v}_{x,y}$  in units of $R_0=h/e^2=2\pi$  for pretty small $F=0.01$ and $\gamma=0.01$. The fact that the approach proves quantization of the Hall conductivity in the case of small $F$ and $\gamma$  is actually not surprising. In fact, in the case of an infinite lattice the standard proof of quantized conductivity involves two steps. The first step is derivation of the Nakano-Kubo equation 
\begin{equation}
\label{d9} 
\sigma_{xy}=\frac{e^2\hbar}{i}\sum_{E_j<E_F<E_{j'}} 
\frac{(v_y)_{j,j'} (v_x)_{j',j} - (v_x)_{j,j'}(v_y)_{j',j}}{(E_j-E_{j'})^2} \;,
\end{equation}
where $(v_x)_{j,j'}$ and $(v_y)_{j,j'}$ are matrix elements of the current operators (\ref{c4}) in the basis of the Landau states $|\Phi_j \rangle$. Notice that, as any linear response result, Eq.~(\ref{d9}) implicitly assumes the limits $F\rightarrow 0$ and $\gamma \rightarrow 0$. In the second step one evaluates this equation by using topological properties of the Landau states \cite{Kohm85}.  The master equation approach simply merges these two steps in one:  we solve the equation for the stationary density matrix $\bar{\rho}$ in the basis of the Landau-Stark states for arbitrary $F$ and $\gamma$ and then take the limits $F\rightarrow 0$ and $\gamma \rightarrow 0$.

We return to the question what fraction of fermionic carriers contribute to the current. While the Bloch states and Landau states pictures do not provide a definite answer to this question, the Landau-Stark states picture indicates that {\em all} carriers move in the $x$ direction with the drift velocity (\ref{d5}) and, hence, contribute to the current. As it was mentioned in Sec.~\ref{sec3b}, the drift velocity $v^*$ is a property of the Landau-Stark states in the limit of small $F$ and $\alpha$. Since the drift velocity is a smooth function of $\alpha$, the steps in the Hall current are exclusively due to the step-like behavior of the number of curries ${\cal N}={\cal N}(\alpha,E_F)$ entering Eq.~(\ref{c5}) for the equilibrium density matrix.  

\subsection{Alignment effects}

We have shown that our approach based on the master equation (\ref{1}) with the relaxation term (\ref{3}) well reproduces the known results of the linear response theory. The main advantage of the approach, however, is its ability to describe the transport beyond the linear response regime. This allows us to address some new effects. One of these effects -- the negative differential conductivity -- was briefly discussed in the previous sections. In this subsection we analyze another interesting effect -- dependence of the Hall current on orientation of the static force relative to the primary axes of the lattice. In what follows we characterize this orientation either by the parameter $\beta=F_x/F_y$ or by the angle $\theta=\arctan(F_x/F_y)$.

If  $\beta$ is a rational number, the stationary current can be calculated by using the algebraic method. Principal possibility for generalization of the algebraic method to arbitrary rational $\beta$ follows from the fact that, similar to the case $\beta=0$, the Landau-Stark states for any rational $\beta$ are Bloch-like states with the band energy spectrum arranged into a ladder. The procedure of calculating the Landau-Stark states for $\beta=r/q$ is described in detail in Ref.~\cite{85}. Unfortunately, this procedure becomes more and more involved when $r$ and $q$ are increased. Besides this, it pre-excludes the case of irrational $\beta$ that is of large theoretical interest. It was proved in Ref.~\cite{90} that for irrational $\beta$ the Landau-Stark states are localized states with the discrete energy spectrum \cite{remark2}. Thus, strictly speaking, the ballistic transport in the system is prohibited if $\beta\ne r/q$. It is not clear {\em a priori} whether this fundamental difference between rational and irrational directions of the static force affects the diffusive transport.
\begin{figure}
\center
\includegraphics[width=12cm, clip]{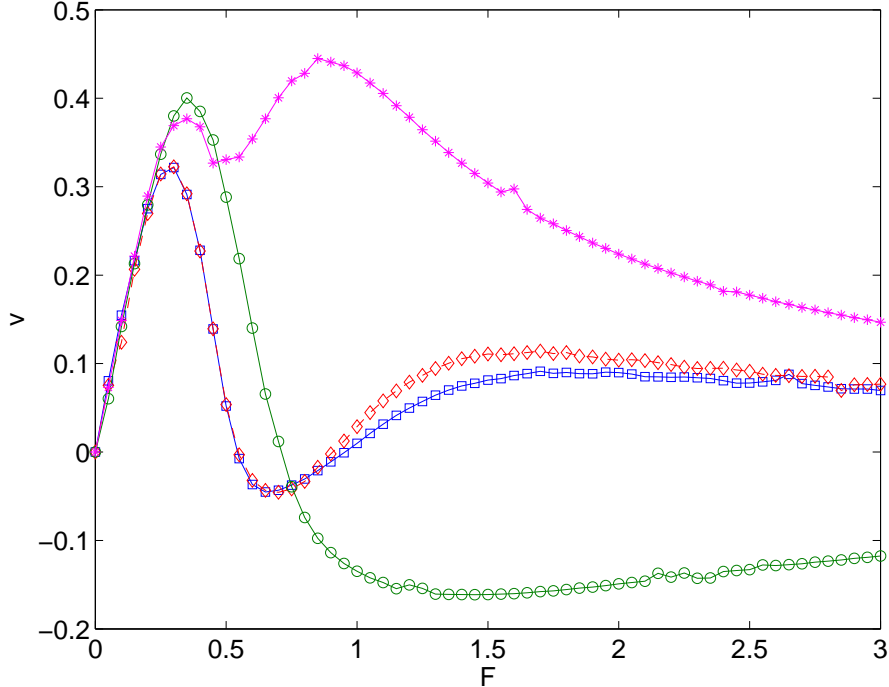}
\caption{Diffusive Hall current for four different orientations of the static force. Parameters are $J_x=J_y=1$, $\alpha=0.1$, $\gamma=0.1$, and  $\beta=0$ (asterisks), $\beta=1/3$ (squares), $\beta=(\sqrt{5}-1)/4\approx1/3$ (diamonds), and $\beta=1$ (circles).} 
\label{fig9}
\end{figure}

To answer the above question we employed the dynamical approach of Sec.~\ref{sec2}.  Notice that now, in the case of two-dimensional lattices, each density matrix element has four indexes that imposes severe limitation on the system size. We tested convergence of the method against exact results depicted in Fig.~\ref{fig6}.  It was found that one can tolerate the error due to finite system size if  $L$ exceeds 20 lattice sites, see asterisks and circles in Fig.~\ref{fig6}.

The stationary current calculated by using the dynamical method is shown in Fig.~\ref{fig9} for the three rational $\beta=0,1/3,1$ (solid lines) and one irrational $\beta=(\sqrt{5}-1)/4\approx 1/3$ (dashed line). Only the Hall currents $\bar{v}_\perp=\bar{v}_x\cos\theta - \bar{v}_y\sin\theta$ are depicted. The corresponding curves for the Ohm current  $\bar{v}_\parallel=\bar{v}_x\sin\theta + \bar{v}_y\cos\theta$ were found to closely follow the red line in Fig.~\ref{fig6} and are not shown.  First of all we notice that the cases of rational $\beta=1/3$ and irrational $\beta=(\sqrt{5}-1)/4$ look similar. Thus, unlike the ballistic transport ($\gamma=0$), the diffusive transport is not sensitive to rationality of the parameter $\beta$. Two important conclusions immediately follow from this result: (i) The limit $\gamma\rightarrow0$ is singular an should be taken with precaution.  Namely,  one should first find the stationary matrix and then take the limit; (ii) To study the stationary current we can restrict ourselves by considering only rational $\beta$, where we can employ the algebraic approach. Using this method one can treat essentially lager lattices (hundreds sites in one directions) to eliminate finite size effects.  Besides this the algebraic method insures the right sequence for taking the limit $\gamma\rightarrow 0$, thus relating ballistic transport to the diffusive transport. 
\begin{figure}
\center
\includegraphics[width=12cm, clip]{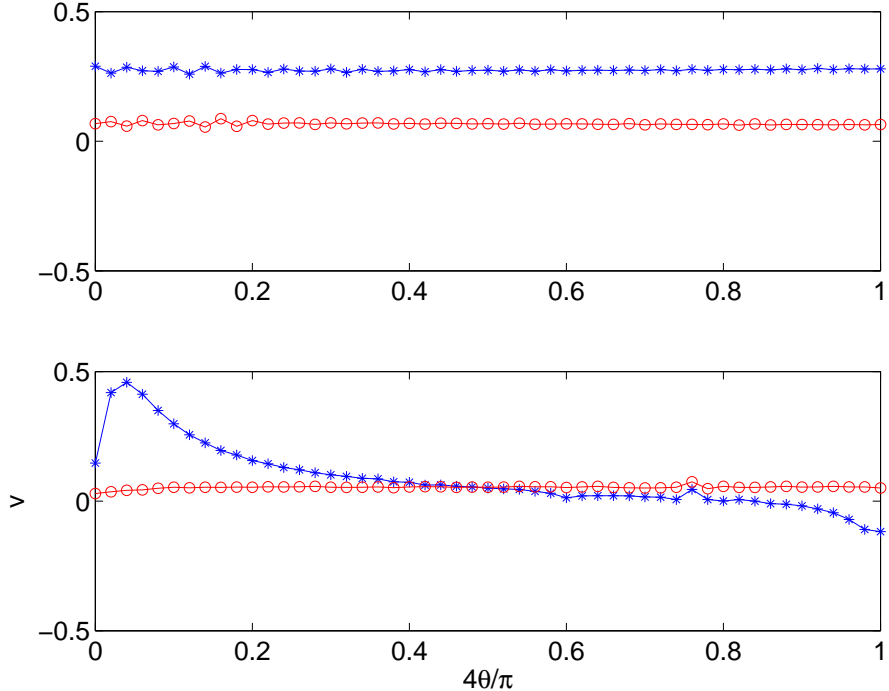}
\caption{Angular dependence of the Hall (asterisks) and Ohm (cirles) currents  for $F=0.25$ (upper panel) and $F=3$ (lower panel). The other parameters are $J_x=J_y=1$, $\alpha=0.1$, and $\gamma=0.1$.} 
\label{fig10}
\end{figure}

Let us discussed numerical results depicted in Fig.~\ref{fig9} in some more detail. In the linear regime the Hall current  $\bar{v}_\perp$ is seen to be independent of the orientation of the static force, which is consistent with the effective mass approximation. However, if $F$ is increased we observe a pronounced dependence of the Hall current on $\theta$, see Fig.~\ref{fig10}. Moreover, for $\theta$ close to $\pi/4$ the Hall current is inverted. 

\section{Conclusions}

We revisited the theory of diffusive transport in the context of cold atoms subject to artificial electric and magnetic fields. The main difference of this system from its solid-state equivalent is a limited applicability of the linear response theory. In this work we treated the problem for arbitrary magnitudes of the electric and magnetic fields. In this sense the only limitation of the presented theory is validity of the tight-binding Hamiltonians that requires negligible Landau-Zener tunneling \cite{94}.

First we considered degenerate bosonic or fermionic carriers in a one-dimensional lattice subject to `an electric' field.  Stationary distributions of  the carries over quasimomentum Bloch states were found as functions of the electric field magnitude. Using these distributions we calculated the diffusive current that was proved to obey the Esaki-Tsu equation \cite{Esak70} with the prefactor depending on the carriers statistics.

Next we considered the fermionic carriers in a  square two-dimensional lattice subject to in-plane `electric' field and  normal  the lattice plane `magnetic' field. In this system one has to distinguish between two currents -- the Ohm current, that flows in the direction parallel to the vector ${\bf F}$ of the electric field, and the Hall current in the direction perpendicular to ${\bf F}$. In the limit of small $F$ our results were shown to reproduce those of the linear response theory,  including the integer quantum Hall effect. The new results refer to the diffusive current beyond the linear response regime. In particular, we found strong dependence of the Hall current on the orientation of the vector ${\bf F}$ relative to the primary axes of the lattice.

In the present work we restricted ourselves by the case of uniform magnetic field and small Peierls's phase $|\alpha|\ll 1/2$  where the system shows some universal features, for example, the Hall current is related to the classical drift velocity.  In our forthcoming publications we will extend the discussed approaches to study diffusive current of cold atoms for $\alpha\sim 1/2$ and the other magnetic field configurations, like the staggered field realized in the recent experiment \cite{Aide11}. The other direction of research is conductance of a finite system that is believed to be determined by the edge states \cite{remark3}.

The authors express his gratitude to D.~N.~Maksimov for useful remarks and acknowledge financial support of Russian Academy of Sciences through the SB RAS integration Project 
No. 29 (Dynamics of atomic Bose-Einstein condensates in optical lattices).



\begin{thebibliography}{99}
\bibitem{remark2}
The spectrum of the localized Landau-Stark states is given by the sum of two Wannier-Stark ladders, i.e, $E_{n,k}=F_x n + F_y k$. This is a consequence of the translational property of the localized Landau-Stark states. Namely, if  $|\Psi\rangle=\sum_{l,m}\psi_{l,m}|l,m\rangle$ is a localized eigenstate of (\ref{c1}) with the energy $E$, then the state 
$|\tilde{\Psi}\rangle=\sum_{l,m} \psi_{l-n,m-k}e^{-i2\pi\alpha kl} |l,m\rangle$
is also an eigenstate with the energy $\tilde{E}=E+(F_x n+F_y k)$.
\bibitem{remark3}
With respect to cold atoms in optical lattices the edge states are discussed in Refs.~\cite{Buch12,Gold12,Gold13}.
\end{thebibliography}
\end{document}